\documentstyle[preprint,aps,epsf,axodraw,12pt]{revtex}
\begin{document}
\tightenlines \preprint{\vbox {
\hbox{hep-th/0311219v2}}}
\title{Harmonic scaling laws and underlying structures}
\author{Ji-Feng Yang}
\address{Department of Physics, East China Normal University,
Shanghai 200062, China}
\maketitle
\begin{abstract}
Based on the effective field theory philosophy, a universal form
of the scaling laws could be easily derived with the scaling
anomalies naturally clarified as the decoupling effects of
underlying physics. In the novel framework, the conventional
renormalization group equations and Callan-Symanzik equations
could be reproduced as special cases and a number of important and
difficult issues around them could be clarified. The underlying
theory point of view could envisage a harmonic scaling law that
help to fix the form of the loop amplitudes through anomalies, and
the heavy field decoupling can be incorporated in this underlying
theory approach in a more unified manner.
\end{abstract}
\vspace{2.0cm}
\section{Introduction}
According to the effective field theory (EFT) point of
view\cite{PWQFTbook}, the insensitivity of the 'low energy'
physics to 'high energy' details allows us to parametrize the
ill-definedness in EFTs in artificial regularization schemes and
then remove them through renormalization program\cite{BPHZ}. The
necessary presence of renormalization procedure and the
insensitivity of the effective theories to the underlying one
leads to the renormalization group equation
(RGE)\cite{SP,GML,HV,WRGE} and anomalous scaling laws encoded in
the Callan-Symanzik equations (CSE)\cite{CS,WRGE}. But all these
equations are prescriptions dependent, and therefore suffer from
various shortcomings associated with the prescriptions. The
prescription dependence problem might be severe in nonperturbative
contexts\cite{PRD65yang,LENU}. Even in perturbative regime it is
in fact not a simple task to remove the prescription dependence,
especially in QCD processes\cite{scheme}. Moreover, the
interpretation of RGE as independence of renormalization points
seems unsatisfactory.

In this report we wish perform a pedagogical but general
derivation of the EFT scaling laws based on existence of
underlying structures to clarify a number of important issues
around RGE and CSE and to uncover the physical significance behind
the prescription dependence. We first introduce low energy
expansion for operators with underlying structures in section two
as a warm up. In section three, we perform the formal derivation
of the scaling laws for vertex functions in any EFT
(renormalizable or not) by only assuming that their underlying
theory definitions exists. The emergence of RGE and CSE and the
problems associated with prescriptions will be discussed in
section four. Section five will be devoted to a 'harmonic'
understanding of the scaling anomalies and a novel use of the
scaling laws illustrated with some simple examples. The decoupling
issue is discussed in the underlying theory perspective in section
six. The whole report will be summarized in section seven.
\section{low energy expansion of operators with underlying
structures} It is not difficult to convince ourselves that there
should general nontrivial structures\footnote{The true underlying
structures remain unknown. But for our purpose, it is enough to
postulate that they exist and make the extremely short distances
physics well defined.} underlying the simple point-particle fields
in EFTs. Therefore, for each EFT operator (elementary or not)
there should be a corresponding one with nontrivial underlying
structures. In long distances or at low energy scales (the
decoupling limit), the underlying structures are effectively
'invisible' and the operators reduce (via expansion) to the EFT
ones for point particles. But in very short distances or at
extremely high energy scales, such underlying structures should
become 'visible' and dictate the short distance dynamics.

Using operator and Hilbert state vector formalism, the above
scenario means the following expansion equations:
\begin{eqnarray} \label{UTOPE1}
&&\widehat{O}([x],[g];\{\sigma\})|LES\rangle
=\widehat{O}([x],[g])|LES\rangle
+{\mathcal{O}}(1/\Lambda_{\{\sigma\}});\\
\label{UTOPE2} && {\prod}_i
\widehat{O}_i([x_i],[g];\{\sigma\})|LES\rangle = {\prod}_i
\widehat{O}_i([x_i],[g])|LES\rangle+
{\mathcal{O}}(1/\Lambda_{\{\sigma\}});\\
\label{UTdecoupling} && \widehat{O}([x];\{\sigma\})|LES\rangle
=\sum_i c_i ([g]) {\widehat{O}}^{\prime}_i([x,[g])|LES\rangle
+{\mathcal{O}}(1/\Lambda_{\{\sigma\}}),
\end{eqnarray}where $x$, $[g]$ and ${\widehat{O}}_j$ denote
respectively the spacetime coordinates (we do not address quantum
gravity for simplicity), the EFT parameters and operators, while
$|LES\rangle$ refers to a generic EFT state and
$\Lambda_{\{\sigma\}}$ refers to the smallest scale among the
underlying parameters $\{\sigma\}$. The first two equations could
be viewed as a generalization of the OPE a la Wilson\cite{Wilson}
and the third one as Witten's heavy quark decoupling\cite{Witten}.

However, for the states of extremely high energy scales, the above
three equations naturally break down, then we must employ the
operators {\em with} underlying structures, otherwise we get UV
ill-definedness. This is what we usually met in Feynman
amplitudes: the intermediate (free particle) states summation
(loop momentum integration) extends to infinite regions where we
must employ nontrivial underlying structures to make the loop
integration finite. Only after the loop integrations are safely
carried out could we perform the expansion with respect to the
underlying scales, that is , the decoupling limit operation and
EFT loop integration generally do not commute. For each loop that
is originally ill defined in EFTs, such limit operation performed
after integration should lead to a definite nonlocal component
{\em and} a finite local component. The coefficients in the local
component would contain additional constants arising from the
decoupling limit that should be definite. In conventional
renormalization prescriptions these constants would be replaced by
various prescription dependent cutoff or subtraction scales. For
the following derivation, we only need that the underlying
structures and hence the finite constants from the decoupling
limit exist in 'right' places. The good thing is we only need
their existence.
\section{Canonical scaling with underlying structures}
Now let us consider a general complete vertex function (1PI)
$\Gamma^{(n)}([p],[g];\{\sigma\})$ that is well defined in the
underlying theory with $[p],[g]$ denoting the external momenta and
the Lagrangian couplings (including masses) in a EFT and
$\{\sigma\}$ denoting the underlying parameters or constants. Now
it is easy to see that such a vertex function must be a
homogeneous function of all its dimensional arguments, that is
\begin{eqnarray}
\label{scaling1} \Gamma^{(n)}([\lambda p],[\lambda^{d_g} g]; \{
\lambda^{d_{\sigma}} \sigma\})= \lambda^{d_{\Gamma^{(n)}}}
\Gamma^{(n)} ([p],[g];\{\sigma\})
\end{eqnarray} where $d_{\cdots}$ refers to the canonical mass
dimension of any parameters involved.

In the decoupling limit ($L_{\{\sigma
\}}\equiv\lim_{\Lambda_{\{\sigma\}}\rightarrow \infty}$) that the
underlying structures look vanishingly small, there will
necessarily arise finite constants $\{\bar{c}\}$\cite{TOE} besides
the EFT couplings and masses, then Eq.(~\ref{scaling1}) becomes
\begin{eqnarray}
\label{scaling2} &&\Gamma ^{(n)}([\lambda p],[\lambda
^{d_g}g];\{\lambda ^{d_{\bar{c}}}{\bar{c}}\})\equiv L_{\{\sigma
\}}\Gamma ^{(n)}([\lambda p],[\lambda
^{d_g}g];\{\lambda ^{d_\sigma }\sigma \})  \nonumber \\
&=&\lambda ^{d_{\Gamma ^{(n)}}}\Gamma
^{(n)}([p],[g];\{{\bar{c}}\})\equiv \lambda ^{d_{\Gamma
^{(n)}}}L_{\{\sigma \}}\Gamma ^{(n)}([p],[g];\{\sigma \}).
\end{eqnarray} Note that $\{{\bar{c}}\}$ only appear in the loop
diagrams or in the pure quantum corrections, as 'agents' for the
hidden participation of the underlying structures. From
Eqs.(~\ref{UTOPE1},~\ref{UTOPE2},~\ref{UTdecoupling}), such
constants could only appear or reside in the local component of
the vertex functions involved, which leads to local
operators\footnote{Mathematically, this could be seen from the
expansion of the finite amplitudes $\Gamma^{(n)}([p],[g];
\{\sigma\})$ in terms of the ratios
$\frac{[p,g]}{\Lambda_{\{\sigma\}}}$ as they tend to zero, which
will give rise to a local polynomial.}. We will further clarify
these constants later.

The differential form for Eq.(~\ref{scaling1}) reads
\begin{eqnarray}
\label{scaling3} \{\lambda \partial _\lambda  + \sum d_g
g\partial_g + \sum d_{\sigma} \sigma \partial_{\sigma}
-d_{\Gamma^{(n)}}\} \Gamma^{(n)}([\lambda p],[g];\{\sigma\})=0.
\end{eqnarray}Noting that the operation $\sum d_g g\partial_g$ inserts
the trace of the stress tensor ($\Theta$) for EFTs in concern,
Eq.(~\ref{scaling3}) could be rewritten as the following
inhomogeneous form,\begin{eqnarray} \label{scaling4} \{\lambda
\partial _\lambda  + \sum d_{\sigma} \sigma
\partial_{\sigma} -d_{\Gamma^{(n)}}\}
\Gamma^{(n)}([p],[g];\{\sigma\}) =i\Gamma^{(n)}_{\Theta}
([0;\lambda p],[g];\{\sigma\}).\end{eqnarray} Obviously the
constants with zero canonical mass dimensions do not contribute to
the scaling behavior. Eq.(~\ref{scaling3}) or Eq.(~\ref{scaling4})
is just the most general underlying theory version of the EFT
scaling laws. The only distinction with naive EFT scaling laws is
the {\bf canonical} scaling contribution from the underlying
structures ($\sum d_{\sigma} \sigma
\partial_{\sigma}$) as the only source for EFT scaling anomalies
in the low energy limit.

To see this point more clearly, let us examine the low energy
limit of the above scaling laws in terms of the agent constants
$\{\bar c\}$ that can be derived from Eq.(~\ref{scaling2}) read
\begin{eqnarray}
\label{scaling5} \{ \lambda \partial _\lambda  + \sum d_{\bar c}
{\bar c}
\partial_{\bar c} +\sum d_g g \partial_g -d_{\Gamma^{(n)}} \}
\Gamma^{(n)}([\lambda p],[g];\{\bar c\})=0.
\end{eqnarray}Comparing this equation with
Eq.(~\ref{scaling3}), we have,
\begin{eqnarray} \label{decoupling} L_{\{\sigma \}}\{\sum
d_{\sigma}\sigma
\partial_{\sigma}\} \Gamma ^{(n)}([p],[g];\{\sigma \})=
 \{ \sum d_{\bar c} {\bar c} \partial_{\bar c}  \}
\Gamma^{(n)}([p],[g];\{\bar c\}),
\end{eqnarray}provided the
underlying structures scenario is valid. We should stress that
Eqs.(~\ref{scaling3},~\ref{scaling4},~\ref{scaling5},~\ref{decoupling})
are valid not only order by order but also graph by graph. Since
the agent constants only appear in the local components of each
loop that is not well defined in EFTs, the variations in these
agent constants naturally induce the insertions of local operators
($I_{O_i}$) in the corresponding loop graphs. Therefore we arrive
at the following decoupling theorem,
\begin{eqnarray}
\label{preRGE}
 &&L_{\{\sigma \}}\{\sum
d_{\sigma}\sigma
\partial_{\sigma}\}=\sum d_{\bar{c}}{\bar{c}}\partial
_{\bar{c}}=\sum_i \delta_{O_i}([g];\{\bar{c}\}) I_{O_i},
\end{eqnarray}where $\delta_{O_i}$ must be functions of the EFT
coupling and masses and the agent constants:
$\delta_{O_i}=\delta_{O_i}([g];\{\bar{c}\})$. That means, in the
decoupling limit of the underlying structures' contribution to the
scaling can be expanded in terms of EFT operators.

As some of these operators appear in EFT Lagrangian with coupling
constants (denoted as $[g]$) as coefficients, then their
insertions can be realized via $\sum_g d_g g\partial_g$. The
kinetic operator insertion for EFT field $\phi $ will be denoted
as ${ \hat{I}}_\phi $. The rest are not present in Lagrangian,
they will be denoted as $O_N$. Therefore, we can write,
\begin{eqnarray}
\label{decomposition} &&\sum_i \delta_{O_i}([g];\{\bar{c}\})
I_{O_i}=\sum_{g}\delta _g([g];\{\bar{c} \})g\partial
_g+\sum_{\phi}\delta _\phi ([g];\{\bar{c}\}){\hat{I}}_\phi
+\sum_{O_N}\delta _{O_N}([g];\{ \bar{c}\}){\hat{I}}_{O_N},
\end{eqnarray}
Now with Eqs.(~\ref{decoupling},~\ref{decomposition}) we can turn
the decoupling theorem in Eq.(~\ref{preRGE}) and the full scaling
of Eq.(~\ref{scaling5}) into the following forms,
\begin{eqnarray}
\label{preRGE1}&&\{\sum_{\bar c} d_{\bar{c}}{\bar{c}}\partial
_{\bar{c}}-\sum_{O_N}\delta _{O_N}{\hat{I}} _{O_N}-\sum_{g}\delta
_gg\partial_g -\sum_{\phi}\delta _\phi { \hat{I}}_\phi\}
\Gamma ^{(n)}([\lambda p],[g];\{\bar{c}\})=0,\\
\label{scaling11} &&\{\lambda \partial _\lambda +\sum_{O_N}\delta
_{O_N}{\hat{I}} _{O_N}+\sum_{g}(d_g+\delta _g)g\partial
_g+\sum_{\phi}\delta _\phi { \hat{I}}_\phi -d_{\Gamma
^{(n)}}\}\Gamma ^{(n)}([\lambda p],[g];\{\bar{c}\})=0.
\end{eqnarray}Here Eqs.(~\ref{preRGE1},~\ref{scaling11}) are only
true for the complete sum of all graphs (or up to a certain
order). It is obvious that $\delta_{O_i}$ are just the anomalous
dimensions and comprise all the scaling anomalies, which just come
from the decoupling limit of the {\bf canonical} scaling behaviors
of the underlying structures, as stated in
Eqs.(~\ref{decoupling}). These two general equations can describe
both the elementary vertex functions' scaling and the ones for
composite operators.

In the underlying theory,
Eqs.(~\ref{scaling3},~\ref{scaling4},~\ref{scaling5},~\ref{preRGE1})
and (~\ref{scaling11}) must be identities. But in practice, the
underlying structures and hence the agent constants
($\{\bar{c}\}$) are unknown. Then Eq.(~\ref{scaling5}) or
(~\ref{scaling11}) could serve as the constraints for these
constants. Later, we will illustrate this point on some simple
one-loop amplitudes in section five.
\section{Novel perspective of RGE and CSE}
In this section we limit our attention to a special
type of theories: the one without the scaling anomalies
$\sum_{\{O_N\}} \delta_{O_N}([g];\{\bar c\}) { \hat{I}}_{O_N}$,
i.e., the renormalizable theories in conventional terminology.
Without these operators, Eqs. (~\ref{preRGE1}) and
(~\ref{scaling11}) become simpler,
\begin{eqnarray}
\label{preRGE2} &&\{\sum_{\bar{c}}d_{\bar{c}}\bar{c}\partial
_{\bar{c}}-\sum_{g}\delta _g g\partial _g-\sum_{\phi }\delta _\phi
{\hat{I}}_\phi \}
\Gamma ^{(n)}([\lambda p],[g];\{\bar{c}\})=0; \\
\label{scaling13} &&\{\lambda \partial _\lambda
+\sum_{\bar{c}}d_{\bar{c}}\bar{c}\partial _{\bar{c}} +\sum_{g}
d_gg\partial _g-d_{\Gamma ^{(n)}}\}\Gamma ^{(n)}([\lambda
p],[g];\{\bar{c}\})\nonumber \\
=&&\{\lambda \partial _\lambda +\sum_{g} (d_g+\delta _g)g\partial
_g+\sum_{\phi} \delta _\phi {\hat{I}}_\phi -d_{\Gamma
^{(n)}}\}\Gamma ^{(n)}([\lambda p],[g];\{\bar{c}\})=0.
\end{eqnarray}
These are just the underlying theory versions for RGE and CSE for
renormalizable theories. The significant differences between the
underlying theory versions and the original ones will be discussed
later in this section. To facilitate the comparisons we will turn
these equations into more familiar forms. For this purpose, we
note that all the agent constants could be parametrized in terms
of a single scale $\bar{\mu}$ as $\{\bar{\mu};({\bar c}_0)\}$ with
each ${\bar c}_0$ ($\equiv \frac{\bar c}{\bar \mu}$) being
dimensionless. In conventional renormalization programs, they are
first predetermined through renormalization conditions, finally
transformed into the physical parameters\cite{sterman} or
optimized somehow\cite{scheme,maxwell}.

We should stress the most striking difference between $[g]$ and
$\{\bar c\}$: The EFT parameters $[g]$ or the tree vertices alone
can not make EFTs well defined. There must be underlying
structures or their agents $\{\bar c\}$ for characterizing the
quantum "paths" of EFTs.
\subsection{RGE and CSE as decoupling theorems}
Mathematically, an 'invariance' for the complete vertex functions
is encoded in the decoupling theorem of Eq.(~\ref{preRGE2}): the
finite variation in the agent constants could be absorbed by the
EFT parameters. This amounts to a finite 'renormalization'
invariance of the EFTs, i.e., the renormalization 'group'. But
this invariance is only valid provided: (1) the finite
renormalization originates from a homogeneous scaling of all the
dimensional parameters; (2) after transformation no scale is
comparable to the underlying scales, otherwise it breaks down as a
decoupling theorem. This is just the underlying theory version of
RGE with nontrivial validity conditions. These premises seem to be
overlooked conventionally.

In Eq.(~\ref{preRGE2}) and (~\ref{scaling13}) the insertion of
kinetic operators appears unfamiliar. To remove it let us make use
of the well known facts that the kinetic insertion only rescales
the line momenta in the graphs: for the fermions, this is
$p\!\!\!/ \rightarrow (1+c_{\psi}) p\!\!\!/$; for bosons, $p^2
\rightarrow (1+c_{\phi})p^2$. Since vertices are joined by lines
(flows of momenta), each line's rescaling in turn leads to the
rescaling of the vertices pair at the two ends of the line. Thus a
vertex is typically rescaled as $\prod_i
(1+c_{\psi_i})^{-1/2}\prod_j(1+c_{\phi_j})^{-1/2}$. For an
$n$-point 1PI vertex function there must be $n$ external momenta
that are not subject to the rescaling, thus to compensate this an
overall rescaling of the complete $n$-point vertex functions must
be introduced. Thus $\sum_{\phi }\delta _\phi {\hat{I}}_\phi$ lead
to the following consequences:
\begin{eqnarray}
\label{scaling16} \delta_g \rightarrow  {\bar \delta}_g\equiv
(\delta_g- n_{g;\phi}\frac{\delta_{\phi}}{2}-n_{g;\psi}
\frac{\delta_{\psi}}{2}),\ \ \ \ \Gamma^{(n_{\phi},n_{\psi})}
\rightarrow (1+c_{\psi})^{n_{\psi}/2}(1+c_{\phi})^{n_{\phi}/2}
\Gamma^{(n_{\phi},n_{\psi})},
\end{eqnarray}with $n_{g;\phi}$ and $n_{g;\psi}$ being
respectively the number of bosonic and fermionic field operators
contained in the vertex with coupling $g$. Then the above
equations take the following forms:
\begin{eqnarray}
\label{RGE} & &\{{\bar{\mu}}\partial_{\bar \mu} - \sum_{g} {\bar
\delta}_g g\partial_g - \sum_{\phi}n_{\phi}
\frac{\delta_{\phi}}{2}- \sum_{\psi}n_{\psi}\frac{
\delta_{\psi}}{2} \} \Gamma^{(n_{\phi},n_{\psi})}
([p],[g];\{\bar{\mu};(\bar{c}^i_0)\})=0;\\
\label{scaling18} & &\{ \lambda \partial_{\lambda}+ \sum_{g} D_gg
\partial_g + \sum_{\phi}n_{\phi}\frac{\delta_{\phi}}{2} +
\sum_{\psi} n_{\psi}\frac{\delta_{\psi}}{2}
-d_{\Gamma^{(n_{\phi},n_{\psi})}}\} \Gamma^{(n_{\phi},n_{\psi})}
([\lambda p],[g];\{\bar{\mu};(\bar{c}^i_0)\}) =0,
\end{eqnarray} with $D_g \equiv {\bar
\delta}_g +d_g$. Now Eqs.(~\ref{RGE}) and (~\ref{scaling18}) take
the familiar forms. One might take the $[g]$ as our finite "bare"
constants as no infinity subtraction is needed in whole
derivation.

For a concrete example, let us consider QED where the conventional
RGE and CSE in Feynman gauge read respectively\cite{WRGE}
\begin{eqnarray}
\label{scaling19} & &\{\mu\partial_{\mu} -\gamma_{m_R} m_R
\partial_{m_R} +\beta
\partial_{\alpha_R}-n_A\gamma_A
-n_{\psi}\gamma_{\psi}\}\Gamma^{(n_A,n_{\psi})} ([p], m_R,
\alpha_R)=0; \\
\label{scaling20} & &\{ \lambda
\partial_{\lambda}+(1+\gamma_{m_R})m_R
\partial_{m_R}-\beta
\partial_{\alpha_R} -n_A\gamma_A -n_{\psi}\gamma_{\psi}
-d_{\Gamma^{(n_A,n_{\psi})}} \} \Gamma^{(n_A,n_{\psi})}([\lambda
p], m_R,\alpha_R)=0,
\end{eqnarray} in a renormalization prescription. To write down
the new versions we note that in QED there are only three
independent series of agent constants:
$[c_m(m,e;\bar{\mu},\bar{c}_{m,0})]$, $[c_e(m,e;{\bar \mu},{\bar
c}_{e,0})=c_{\psi}(m,e;\bar{\mu},\bar{c}_{\psi,0})]$, and
$[c_A(m,e;\bar{\mu},\bar{c}_{A,0})]$ (for $m\bar{\psi}\psi$,
$e\bar{\psi} A\!\!\!/ \psi$, $i\bar{\psi}
\partial \!\!\!/\psi$ and $\frac{1}{4}F^2$) thanks to gauge
invariance. Then
Eqs.(~\ref{preRGE2},~\ref{scaling13},~\ref{RGE},~\ref{scaling18})
imply the following equations:
\begin{eqnarray}
\label{scaling22}& &\{{\bar \mu}
\partial_{\bar \mu} -\delta_m m\partial_m-\delta_e
e\partial_e-\delta_A {\hat{I}}_A -\delta_{\psi} {\hat{I}}_{\psi}\}
\Gamma^{(n_A,n_{\psi})} ([p], m,e|\{c_m,c_e,c_A,c_{\psi}\})
\nonumber \\
&=&\{ {\bar \mu} \partial_{\bar \mu} - {\bar \delta}_m m\partial_m
-{\bar \delta}_e e\partial_e - n_{\phi}\frac{\delta_{\phi}}{2}-
n_{\psi} \frac{\delta_{\psi}}{2}\}\Gamma^{(n_A,n_{\psi})} ([p],
m,e|\{c_m,c_e,c_A,c_{\psi}\})=0 ; \\
\label{scaling23} & &\{\lambda\partial_{\lambda}+m\partial_m
+\bar{\mu}\partial_{\bar \mu}-d_{\Gamma^{(n_A,n_{ \psi})}}\}
\Gamma^{(n_A,n_{\psi})} ([\lambda p], m,e|
\{c_m,c_e,c_A,c_{\psi}\})\nonumber \\
&=&\{\lambda\partial_{\lambda}+(1+\delta_m)m\partial_m +\delta_e
e\partial_e +\delta_A {\hat{I}}_A+\delta_{\psi}
{\hat{I}}_{\psi}-d_{\Gamma^{(n_A,n_{ \psi})}}\}
\Gamma^{(n_A,n_{\psi})} ([\lambda p], m,e|
\{c_m,c_e,c_A,c_{\psi}\})  \nonumber \\
&=&\{ \lambda \partial_{\lambda}+ (1+{\bar \delta}_m )m
\partial_m +{\bar \delta}_ee\partial_e+ n_A\frac{\delta_A}{2} +
n_{\psi}\frac{\delta_{\psi}}{2} -d_{\Gamma^{(n_A,n_{\psi})}}\}
\nonumber \\
& &\times\Gamma^{(n)}([\lambda p], m,e|
\{c_m,c_e,c_A,c_{\psi}\})=0.
\end{eqnarray} Thus the forms are in conformity after noting the
following correspondence:\begin{eqnarray} \label{scaling24}
\gamma_{m_R} \sim {\bar \delta}_m\equiv (\delta_m-\delta_{\psi})
;\ \ \beta (\alpha_R)/\alpha_R=2\gamma_A \sim \delta_A \sim
-2{\bar \delta}_e;\ \ 2\gamma_{\psi} \sim \delta_{\psi}.
\end{eqnarray} Here we note
again that the first version of scaling law in
Eq.(~\ref{scaling23}) (i.e.,
$\{\lambda\partial_{\lambda}+m\partial_m +\bar{\mu}\partial_{\bar
\mu}-d_{\Gamma^{(n_A,n_{ \psi})}}\} \Gamma^{(n_A,n_{\psi})}=0$,
$\sum d_{\bar c}
\bar{c}\partial_{\bar{c}}=\bar{\mu}\partial_{\bar{\mu}}$ as $\sum
d_{\bar{c}_0} \bar{c}_0\partial_{\bar{c}_0}=0$!) are also valid
order by order and graph by graph. This property could lead to a
useful means for determining the constants to certain degree, see
section five.

Similarly the full scaling laws for the generating functional
read,
\begin{eqnarray}
\label{RGEGen}&&\{\sum_{\bar c} d_{\bar c} {\bar c}
\partial_{\bar c}-\sum_{g}\delta_gg\partial_g
-\sum_{\phi} \delta_{\phi}{\hat{I}}_{\phi}\}  \Gamma^{1PI}
([\phi],[g];\{\bar c\})=0;\\
 \label{ScalingWT} & &\left \{\sum_{\phi}\int d^D x
[d_{\phi}-x\cdot\partial_x) \phi(x)] \frac {\delta}{\delta
\phi(x)} + \sum_{g}d_gg\partial_g +\sum_{\bar c} d_{\bar c} {\bar
c} \partial_{\bar c}-D\right \}  \Gamma^{1PI} ([\phi],[g];\{\bar
c\}) \nonumber \\
&=&\left \{\sum_{\phi}\int d^D x [d_{\phi}-x\cdot\partial_x)
\phi(x)] \frac {\delta}{\delta \phi(x)} + \sum_{g}D_gg\partial_g
+\sum_{\phi} \delta_{\phi}{\hat{I}}_{\phi}-D\right \} \Gamma^{1PI}
([\phi],[g];\{\bar c\}) =0.
\end{eqnarray}
with $D$ denoting the spacetime dimension. Again this is correct
for any consistent EFTs.

We note that the operator trace anomalies could be readily read
from Eq.(~\ref{ScalingWT}). For QED this is simply the following
\begin{eqnarray}
\label{scaling26}g_{\mu \nu }{\Theta }^{\mu \nu }=(1+\delta
_m)m{\bar{\psi}}\psi +\frac {1}{4}\delta _AF^{\mu \nu }F_{\mu \nu
}-\delta _\psi i{\bar{\psi}} D\!\!\!\!/ \psi=(1+{\bar
\delta}_m)m{\bar{\psi}}\psi +\frac 14\delta _AF^{\mu \nu}F_{\mu
\nu }.
\end{eqnarray}
In the last step we have used the motion equation. Taking
Eq.(~\ref{scaling24}) into account, this agrees exactly with that
given in Ref.\cite{Trace} in form. In unrenormalizable theories,
the trace anomalies would contain an infinite sum of local
composite operators.
\subsection{Implications from the underlying structures' perspective}
Now it is time for us to discuss the consequences following from
the general parametrization of the scaling laws derived in the
underlying structures perspective. We will focus on renormalizable
EFTs. To proceed, let us rewrite Eqs.(~\ref{preRGE2}) and
(~\ref{scaling13}) with a variant parametrization of both the EFT
parameters ($[\tilde{g}]$) and the agent constants
($\{{\tilde{c}}\}$),
\begin{eqnarray}
\label{varRGE2}&&\{\sum_{\tilde{c}}
d_{\tilde{c}}{\tilde{c}}\partial_{\tilde{c}}-\sum_{\tilde{g}}\delta
_{\tilde{g}}{\tilde{g}}\partial_{\tilde{g}} -\sum_{\phi}\delta
_\phi { \hat{I}}_\phi\} \Gamma
^{(n)}([p],[\tilde{g}];\{{\tilde{c}}\})=0,\\
\label{varCSE}&&\{\lambda \partial _\lambda +\sum_{\tilde{g}}
d_{\tilde{g}} \tilde{g}\partial _{\tilde{g}}+\sum_{\tilde{c}}
d_{\tilde{c}}{\tilde{c}}\partial_{\tilde{c}} -d_{\Gamma
^{(n)}}\}\Gamma ^{(n)}([\lambda p],[\tilde{g}];\{{\tilde{c}}\})=0.
\end{eqnarray}Mathematically, one can formally reproduce the
RGE and CSE in any renormalization prescription by appropriately
defining $\{{\tilde{c}}\}$ and $[\tilde{g}]$. Here a prescription
is characterized by the parametrization $\{\bar c\}$ or
$\{{\tilde{\mu}};({\tilde{c}}_0)\}$, unlike the conventional
approaches\cite{scheme}. Furthermore, with each of
$\{{\tilde{c}}\}$ taken as mathematically independent (not merely
fixed within a prescription 'orbit'), Eq.(~\ref{varRGE2}) could
{\em also} describe any consistent prescription variations: the
unified form of the St\"ukelberg-Petermann equations\cite{SP,LB}.
Thus, Eqs.(~\ref{varRGE2},~\ref{varCSE}) (or
Eqs.(~\ref{preRGE2},~\ref{scaling13})), are simpler and more
universal than any conventional version of RGE and CSE, with which
we could clarify a number of important and difficult issues around
renormalization prescriptions and the applications of RGE and CSE.

However, as stressed in the beginning of last subsection, the
'invariance' in Eq.(~\ref{varRGE2}) or (\ref{preRGE2}) is valid
only for the complete vertex functions {\em under} two conditions,
i.e., it is a restricted invariance, in contrast to the
conventional convictions. Once truncated, this invariance only
makes sense up to the truncated order, and the unaccounted higher
orders' contribution {\em must} be under control or smaller, which
turns out to be a criterion for renormalization prescription. Then
we could see the possible problems with the conventional
prescriptions: (1) for the full EFT, it is not warranted that an
arbitrary variation of the agent constants could lead to a
prescription that is perfectly equivalent to the original physical
underlying theory parametrization; (2) for truncated series, it is
not true that an arbitrary subtraction prescription could qualify
for the criterion noted above. It is known that certain
prescriptions fail in unstable EFT sectors\cite{Poleren} or are
flawed somehow like IR\cite{Collins} or threshold divergences in
the renormalization constants, so are the RGE and CSE defined
there.

On general grounds, this could be understood as follows: the
physical information is parametrized in the functional dependence
upon {\em all} the domain variables: $([p],[g];\{\bar c\})$,
especially on the physical momenta. All the physical threshold
effects and other momenta behavior of all the EFT observables (or
all the vertex functions) {\em must be preserved or reproduced} in
the variation of the agent constants and the EFT parameters. In
homogeneous rescaling of all the domain variables in a physical
parametrization, the momentum behavior is preserved or
'invariant'. But once a variation (for example, one that is not
homogeneous in all the variables) in the agent constants does
affect the momentum behaviors up to the truncated order, the
perturbative invariance is violated in effect.

We could also view from the reverse angle: since the momenta does
not vary with prescription, then different definition of the EFT
parameters amounts to expanding the theory in terms of different
the EFT parameters. Then we need to ask whether the expansions at
different places of the parameters space are equivalent to each
other. To approach the answer, firstly we note that there could be
troubles like IR singularities mentioned above. Secondly we should
be clear that the true physical parametrization is not done before
confronting with physical data\cite{sterman}, as pure theoretical
definition from the underlying theory is unavailable. Then the
definition in a prescription ('renormalized' couplings and masses)
should not preset any special momentum behavior through
subtractions in momentum space. Otherwise, it might preclude the
later confrontation with physical data or boundaries.

Thus, in the stage of theoretical calculation, any parametrization
of the EFT parameters and agent constants ($[\bar{g}]; \{\bar
c\}$) or ($[\tilde{g}];\{{\tilde{c}}\}$) only takes or should take
symbolic meaning before confronting with physical data or
boundaries. This observation applies to any calculational
programs. From the underlying theory point of view, this
indeterminacy is very natural. Of course in some EFTs, the
'classical' parameters could be readily determined, then problem
is to determine the agent constants\footnote{This is like what we
do in solving the differential equations in quantum mechanics or
classical electrodynamics: to determine the unknown constants in
the solutions, one must impose concrete physical boundary
conditions in terms of given masses and couplings.}.

Now let us remark on a technical aspect of the conventional
subtraction prescriptions and the RGE and CSE defined there: a
number of different disguises for each EFT coupling constant $g$
are involved\cite{Coq}, which are in turn the infinite bare
coupling $g_B$, the renormalized coupling (prescription dependent)
$g_R$, the running coupling $\bar{g}(\lambda)$ due to momentum
rescaling ($p\rightarrow \lambda p$), renormalization point
independent coupling
 $\widehat{g}$ (but still prescription dependent),
the physically determined coupling $g_{\text{phys}}$, and the
effective charge $g_{\text{eff}}(q^2)$ that is in fact a physical
form factor. The former four are either infinite or prescription
dependent, hence unphysical and will not appear in the physical
parametrization after confronting with physical data, at least in
principle. In other words, the subtraction programs seem to
introduce inconveniences or make things unnecessarily complicated.

In the parametrization adopted here, we only need $[g]$ and
$\{\bar c\}$ that could first be taken as unspecified in the stage
of theoretical calculations, then fixed somehow through physical
boundaries or data, at most we could introduce a running one
$[\bar{g}(\lambda)]$ from physical rescaling of $[g]$ based on
Coleman's bacteria analogue\cite{Coleman}, therefore formulation
is significantly simplified. This is because the solution to
Eq.(~\ref{scaling13}) can be found as the solution to the
following equation,
\begin{eqnarray}
\label{bacteria} &&\{\lambda \partial _\lambda +\sum_{\bar g}
[d_{\bar g}+\delta _{\bar g}([\bar g];\{\bar c\})]{\bar g}\partial
_{\bar g}+\sum_{\phi} \delta _\phi ([\bar g];\{\bar c\})
{\hat{I}}_\phi -d_{\Gamma ^{(n)}}\}\Gamma ^{(n)}([\lambda p],[\bar
g];\{\bar{c}\})=0
\end{eqnarray}by introducing a running $\bar g$ for each EFT
parameter $g$ ($d_{\bar g}=d_g$) that satisfies the following kind
of equation,
\begin{eqnarray}
\label{running} \lambda \partial _\lambda \{{\bar
g}([g];\lambda)/\lambda^{d_g}\}=-\{d_{\bar g}+\delta _{\bar
g}([\bar g];\{\bar c\})\}{\bar g}([g];\lambda)/\lambda^{d_g},
\end{eqnarray}with the natural boundary condition: ${\bar
g}([g];\lambda)|_{\lambda=1}=g$ for each EFT parameter. If we take
the undetermined EFT parameters $[g]$ as given by the underlying
theory, then we could take them as finite 'bare' parameters that
are also physical. The {\em finite} 'renormalization' constants
can be defined as $z_g([g];\lambda)\equiv{\bar g}([g];\lambda)/g$.

For other operators (kinetic or composite) that have nonzero
anomalous dimensions, say $\hat{O}_i$, we could first introduce
for each of them a fictitious coupling, say $C_i$, then the finite
renormalization constants could be defined as
$z_{\hat{O}_i}\equiv{\bar C}_i (\lambda)/C_i$. Similarly we could
introduce the finite 'renormalization' constant matrix for
'mixing' operators\cite{opmixing}. The finite 'renormalization'
constants thus obtained for the field operator could naturally
satisfy the constraint imposed by the K\"allen-Lehmann spectral
representation. As a result, in the underlying theory point of
view, the 'renormalization' constants are finite and could be
introduced afterwards as byproducts, not as compulsory components.
Therefore no complications associated with the infinities
manipulations are needed. We suspect that this scenario might be
very helpful in the complicated EFTs, e.g., the Standard model,
especially in its unstable sector and the CKM matrix. Again we
stress that the running should {\em not} be extrapolated to
extremely high scales where the underlying structures' decoupling
might fail, an important warning that is usually overlooked in the
applications of the conventional RGE or CSE, as no underlying
structures are conceptually incorporated there.

By now we have not been specific in how to obtain the finite loop
amplitudes except the existence of the underlying structures that
is employed as a postulate. In next section, we will briefly
explain a differential equation approach for computing loop
amplitudes based on the underlying structures' scenario\cite{TOE}.
\section{Constraints on the agent constants: harmony notion}
In the preceding sections, we have seen that the EFT scaling
anomalies originate in fact from the normal scaling laws of the
indispensable underlying structures. Therefore taking the
underlying structures into account there is a perfect harmony in
the scaling of everything: the typical EFT parameters and the
typical underlying parameters or the agent constants for the
latter. This is the harmony notion that is natural only in the
presence of the underlying structures. In the scaling of
everything, as the underlying structures are not directly
observable, their normal contributions (via the agent constants)
will therefore be exposed in terms of the low energy variables as
anomalies. Since the low energy dynamics (including the anomalies)
should be definite over distances large enough than the underlying
structures, then from homogeneity or harmony, the agent constants'
scaling are locked with the definite EFT scaling anomalies.
Conversely, this locking from harmony could serve as constraints
on the agent constants as they are unknown in practice, just like
gauge invariance helps to constrain the agent constants before
confronting with physical data or before the true underlying
structures are discovered. Similar understanding of anomalies from
both EFT and underlying theory perspectives is familiar in chiral
anomaly\cite{Hooft}. We hope this harmony principle could be
generalized to other transformation behaviors of EFTs.
\subsection{Scaling anomalies and agent constants: simple examples}
First let us explain our point on two very simple 1-loop vertices
in massless QED: electron self-energy $\Sigma ^{(1)}$ and vacuum
polarization $\Pi ^{(1)}_{\mu \nu }$. One can use any
regularization to compute the definite part and just leave the
local part ambiguous, or use the differential equation method
adopted in Ref.\cite{TOE} based on the existence of nontrivial
underlying stuctures\footnote{The use pioneering use of the
differential equation method could be found in Ref.\cite{Diffr}.
The author is very grateful to the referee for this important
information. This method proves powerful in dealing with the
notorious overlapping divergences\cite{Caswell}.} that will be
explained later,
\begin{eqnarray}
&&\Sigma ^{(1)}(p,-p)=-i\frac{e^2p\!\!\!/}{16\pi ^2}\ln
\frac{-p^2}{{\bar{c}^2_{\psi;(1)}}},\ \ \ \Pi ^{(1)}_{\mu \nu
}(p,-p)=i\frac{e^2}{12\pi ^2}(g_{\mu \nu}p^2-p_\mu p_\nu )\ln
\frac{-p^2}{{\bar{c}}_{A;(1)}^2}.
\end{eqnarray}Here all the agent constants are obvious. These
objects have scaling anomalies,
\begin{eqnarray}
&&(p\cdot \partial _p-1)\Sigma ^{(1)}=-i\frac{e^2}{8\pi
^2}p\!\!\!/,\ \ \ \ \ \ (p\cdot \partial _p-2)\Pi ^{(1)}_{\mu \nu
}=i\frac{e^2}{6\pi ^2}(p^2g_{\mu \nu }-p_\mu p_\nu ).
\end{eqnarray}Then, if the agent constants'
contributions are included, the normal scaling behavior or exact
scale harmony in Eq.(~\ref{scaling5}) or (~\ref{scaling13}) is
restored order by order, graph by graph:
\begin{eqnarray}
\{p\cdot \partial _p+{\bar{c}_{\psi;(1)}}\partial
_{\bar{c}_{\psi;(1)}}-1\}\Sigma ^{(1)}=0,\ \ \ \ \ \ \{p\cdot
\partial _p+{\bar{c}_{A;(1)}}\partial _{\bar{c}_{A;(1)}}-2\}\Pi
^{(1)}_{\mu \nu }=0.
\end{eqnarray}
This is just harmony we have explained above. Of course the agent
constants' scaling becomes EFT scaling anomalies once expressed in
terms of EFT parameters,
\begin{eqnarray}
{\bar{c}_{\psi;(1)}}\partial _{\bar{c}_{\psi;(1)}}\Sigma
^{(1)}=i\frac{e^2}{8\pi ^2}p\!\!\!/,\ \ \ \ \ \
{\bar{c}_{A;(1)}}\partial _{\bar{c}_{A;(1)}}\Pi ^{(1)}_{\mu \nu
}=-i\frac{e^2}{6\pi ^2}(p^2g_{\mu \nu }-p_\mu p_\nu ).
\end{eqnarray} Thus the agent constants are locked by the
scaling anomalies and must appear as $\ln
\frac{1}{{\bar{c}}_{\cdots}^2}$ to balance the dimension in $\ln
p^2$ in the definite nonlocal component. In the operator insertion
version (only valid for the complete or truncated vertices), the
scaling laws take the following forms up to 1-loop level,
\begin{eqnarray}
& &\{p\cdot\partial_p+\delta_{\psi}^{(1)}{\hat{I}}_{\psi}-1\}
\Gamma^{(1)}_{\psi}=0,\ \ \ \ \ \
\{p\cdot\partial_p+\delta_A^{(1)} {\hat{I}}_A-2\}
\Gamma^{(1)}_{\mu\nu}=0,
\end{eqnarray}with the anomalous dimensions $\delta^{(1)}_{\psi}=
\frac{e^2}{8\pi^2},\delta^{(1)}_A =\frac{e^2}{6\pi^2}$.

The anomalies are independent of mass. To see this we can
recalculate the vertices considered above in the massive case.
Here we just discuss the vacuum polarization following the
parametrization considered by Chanowitz and Ellis\cite{Chanow},
the only agent constant appears in the local component,
\begin{eqnarray}
\label{PI} & &\Pi^{\mu\nu}=-\frac{ie^2}{12\pi^2}(p^{\mu}p^{\nu}-
g^{\mu\nu}p^2)\left \{C (m;{\bar c}_A)+6\int^1_0dzz(1-z)
\ln[1-\frac{z(1-z)p^2}{m^2}]\right \}; \\ &
&\Delta^{\mu\nu}=\frac{e^2}{\pi^2} (p^{\mu}p^{\nu}-g^{\mu\nu}p^2)
\frac{m^2}{p^2}\left [\frac{2m^2}{p^2\tau}\ln \frac{ 1-\tau}
{1+\tau}-1\right ], \ \ \tau\equiv\sqrt{1-4m^2/p^2},
\end{eqnarray} with $\Delta^{\mu\nu}$ being obtained through the
canonical trace operator insertion, or equivalently by
$i(m\partial_m)\Pi^{\mu\nu}$. The anomalous scaling equation for
the vacuum polarization reads,\begin{eqnarray}\label{tracei_d} &
&\{p\cdot\partial_p-2\}\Pi^{\mu\nu}=
i\Delta^{\mu\nu}+i\frac{e^2}{6\pi^2}
(g^{\mu\nu}p^2-p^{\mu}p^{\nu}),
\end{eqnarray} where as anticipated the anomaly is exactly the
same as in the massless case. The agent constant appears in the
unknown constant $C (m;{\bar c}_A)$ which is also an unknown
function of mass. Let us determine its functional form using the
harmony notion.

To this end, we note that Eq.(~\ref{tracei_d}) could be put into
the following form as the vacuum polarization tensor must be
homogenous functions in all dimensional parameters, $p,m,{\bar
c}_A$:
\begin{eqnarray}
\label{homo} \{p\cdot\partial_p +m\partial_m +{{\bar
c}_A}\partial_{{\bar c}_A}-2\}\Pi^{\mu\nu}=0.
\end{eqnarray}Since $i\Delta^{\mu\nu}=
-m\partial_m\Pi^{\mu\nu}$, then from Eq.(~\ref{PI}) we could find
after some calculation that $-m\partial_m \Pi^{\mu\nu}=
\frac{ie^2}{12\pi^2}(p^{\mu}p^{\nu}- g^{\mu\nu}p^2)[m\partial_m C
(m;{\bar c}_A) -2] +i\Delta^{\mu\nu} $, that is $m\partial_m C
(m;{\bar c}_A) -2=0$. Then combining this fact with
Eq.(~\ref{homo}), we could obtain the following the equation,
\begin{eqnarray}
& &0=(m\partial_m+{{\bar c}_A}\partial_{{\bar c}_A}) C (m;{\bar
c}_A)={{\bar c}_A}\partial_{{\bar c}_A}C (m;{\bar c}_A)+2,
\end{eqnarray}from which we have
\begin{eqnarray}
\label{CA} C (m;{\bar c}_A)=-\ln \frac{{\bar{c}_A} ^2}{m^2}.
\end{eqnarray}From this functional form, we could say that the
agent constant is 'locked' in the full scaling, the harmony
notion. Or each agent constant associated with a ill-defined loop
integration must appear in the place similar to Eq.(~\ref{CA}).

It is easy to see that all the EFT vertices should exhibit the
same feature. This is quite a natural conclusion following from
the underlying theory point of view. In this point of view, any
prescription where the subtraction scale (as an agent constant)
appears in ways that are inequivalent to the way described by
Eq.(~\ref{CA}) seems unnatural. Here we must mention that the
dependence on the agent constants in the way specified in
Eq.(~\ref{CA}) conventionally leads to the failure of heavy
particles decoupling in the running couplings\cite{Appel}, which
is not a serious problem but often costs theoretical and
computational labors\cite{EFT}. We will exclusively discuss this
issue in the underlying structures' perspective in section six.
\subsection{A technical explanation based on differential equation method}
The natural appearance of the dimensional agents in the way
dictated in Eq. (~\ref{CA}) can also be technically understood in
the differential equation method. First we need to briefly review
the method solely based on the existence of the underlying
structures\cite {TOE} as follows: For a 1-loop graph $G$ of
superficial divergence degree $\omega_G-1$ \cite{Weinb} with its
underlying version amplitude denoted as
$\Gamma_G([p],[g];\{\sigma\})$, we have
\begin{eqnarray}
\label{37} & &\partial^{\omega_G}_{[p]} L_{\{\sigma\}}
\Gamma_G([p],[g]; \{\sigma\})= L_{\{\sigma\}}
\partial^{\omega_G}_{[p]} \Gamma_G([p],[g];\{\sigma\})
\nonumber \\
&=&L_{\{\sigma\}} \int d^dk[\partial^{\omega_G}_{[p]}
g([p,k],[g];\{\sigma\})]=\int d^dk[\partial^{\omega_G}_{[p]}
L_{\{\sigma\}}g([p,k],[g];\{\sigma\})]  \nonumber \\
&=&\int d^dk g^{\omega_G}
([p,k],[g];\{\sigma\})=\Gamma_G^{\omega_G}([p],[g]),
\end{eqnarray} where $\partial^{\omega_G}_{[p]}$ denotes the
differentiation for $\omega$ times with respect to the external
momenta and the resulting amplitude $\Gamma_G^{\omega_G}$ is
convergent or definite already in EFT. That means, before knowing
the underlying structures, we could at most determine the true
decoupling limit of the amplitude $\Gamma_G$ up to an ambiguous
local component (as a polynomial with power index $\omega_G-1$ in
EFT parameters, $N^{(\omega_G-1)}([p],[g])$) by performing the
indefinite integration on the definite amplitude
$\Gamma_G^{\omega_G}$ with respect to the external momenta for
$\omega_G $ times:
\begin{eqnarray} &&\Gamma_G([p],[g];\{{\bar c}_{G}\})\equiv
L_{\{\sigma\}} \Gamma_G([p],[g];\{\sigma\})=
[\partial^{\omega_G}_{[p]}]^{-1} \Gamma_G^{\omega_G}([p],[g])
\text{mod}N^{(\omega_G-1)}([p],[g])
\end{eqnarray}
with $[\partial^{\omega_G}_{[p]}]^{-1}$ denotes the indefinite
integration. To us the agent constants $\{\bar {c}_G\}$ are
unknown. The deduction in Eq.(~\ref{37}) is justified by the
existence of the underlying structures make the EFTs well
defined\cite{TOE}.

As the amplitude $\Gamma _G^{\omega _G}([p],[g])$ is finite (with
negative scaling dimension) then it should  take the following
form (under appropriate Feynman parametrization)
\begin{equation}
\Gamma _G^{\omega _G}([p],[g])\sim \int_0^1[dx]f(\{x\})\frac{
{\mathcal{N}}([p],[g]|\{x\})}{{\mathcal{D}}([p],[g]|\{x\})},
\end{equation} where ${\mathcal{N}}$ and ${\mathcal{D}}$ are
respectively the polynomials in terms of external momenta and/or
masses with the help of Feynman parameters $\{x\}$, with the
fraction ${\mathcal{N}}/{\mathcal{D}}$ scales as $1/p$. Then after
the indefinite integration is done a logarithmic factor will
necessarily appear,
\begin{equation}
[\partial^{\omega_G}_{[p]}]^{-1}\Gamma _G^{\omega _G}([p],[g])\sim
\int_0^1[dx]f(\{x\}){\mathcal{P}}([p],[g]|\{x\})\ln \left [
\frac{{\mathcal{Q}}([p],[g]|\{x\})}{\bar
{c}_{\text{scale}}^2}\right ],
\end{equation}with ${\mathcal{P}}$ and ${\mathcal{Q}}$ being again
two polynomials in EFT dimensional parameters, with the latter
being usually quadratic in momenta. The agent constant (the scale
$\bar {c}_{\text{scale}}$) naturally appears to balance the
dimension of ${\mathcal{Q}}$. If one trades the constant scale in
the logarithmic factor with an EFT mass parameter that appear at
least in one of the lines in the loop, say $m$, so that the
nonpolynomial function of momenta only involve EFT parameters,
then the agent scale would only appear in the polynomial part
\begin{eqnarray}\label{largemass}
-\int_0^1[dx]f(\{x\}){\mathcal{P}}([p],[g]|\{x\})\ln \frac{\bar
{c}_{\text{scale}}^2}{m^2}\end{eqnarray} just as we saw from scale
anomaly analysis. The constant factor multiplying this logarithmic
function is the anomalous dimension. This logarithmic factor is
the key to understand heavy field decoupling in the 'running' of
EFT parameters, as will be shown in next section.

Thus we could see from the foregoing general arguments that the
agent parameters are generally dimensional constants or scales,
they appear in most situations in the argument of a logarithmic
function of external momenta and masses to balance the
dimensions\footnote{Some agent constants appear not as balancing
scales, e.g., in the power law divergent loops\cite{PRD65yang}.
Such agent constants' variation do not 'renormalize' the EFT
parameters.}. They should, in perturbative Feynman diagrammatic
language, appear in each loop integration that is ill defined in
EFT.
\subsection{A simple use of the new version CSE: asymptotic freedom}
Before closing this section we demonstrate a simple use of the new
version scaling law: the reproduction of the asymptotic freedom of
QCD\cite{AF}. To this end consider the following scaling law
satisfied by the two-point gluon correlation function (in Feynman
gauge)
\begin{eqnarray}
\label{preAF} \{\lambda\partial_{\lambda} + {\bar \delta}_{g_s}
g_s\partial_{g_s}+ \sum (1+ {\bar \delta}_{m_i})
m_i\partial_{m_i}-\delta_A\}D^c(\lambda p, g_s,[m_i];\{\bar
c\})=0.
\end{eqnarray}
Taking all quarks as massless (high energy region) and introducing
$\alpha= g^2_s/(4\pi)$, from Eq.(~\ref{preAF}) we could obtain the
following equation for the effective coupling
$\alpha_{eff}\equiv\alpha D^c$,
\begin{eqnarray}
\label{Aeff} (\lambda\partial_{\lambda}-\delta_A \alpha
\partial_{\alpha}) \alpha_{eff}(\lambda p, \alpha;\{ \bar
c\})=0.
\end{eqnarray}Here we used that ${\bar \delta}_{g_s}=-\frac{1}{2}
\delta_{A}$. Truncating $\delta_A$ at one loop level we have
$\delta^{(1)}_A \alpha=-\frac{1}{2}l_f\alpha$ ($l_f$ is a flavor
dependent number), and there is only one agent scale, ${ \bar
c}_{A;(1)}$, which allows us to make the following Taylor
expansion,
\begin{eqnarray}
 \alpha^{(1)}_{eff}=\sum_{n=0}^{\infty} f_n (\alpha) t^n,\
\ t\equiv\frac{1}{2}\ln(p^2/ {{ \bar c}_{A;(1)}}^2),
\end{eqnarray} then substitutes it into Eq.(~\ref{Aeff}), one can
find the following relation,
\begin{eqnarray}
\label{45} & &nf_n=\delta^{(1)}_A \alpha\partial_{\alpha} f_{n-1},
\ \ f_0=\alpha,\ \ \ \forall n\geq 1.
\end{eqnarray}
The solution to Eq.(~\ref{45}) and therefore the solution to
Eq.(~\ref{Aeff}) are easy to find as
\begin{eqnarray}
\label{47} &&f_n=a^n\alpha ^{n+1},\ \ \forall n\geq 0,\ \
\Rightarrow\alpha _{eff}^{(1)}(p^2,\alpha
;\bar{c}_{A;(1)}^2)=\frac \alpha {1-\frac 12l_f\alpha \ln
(p^2/{\bar{c}_{A;(1)}^2})}.
\end{eqnarray}For QCD it is well known that $l_f$ is negative and
hence the effective charge in Eq.(~\ref{47}) exhibit asymptotic
freedom. Similar phenomenon could be exhibited by introducing a
running coupling for $\alpha$ as in Eq.(~\ref{running}):
$\lambda\partial_{\lambda}\bar{\alpha}(\lambda)-
\delta_A^{(1)}(\bar{\alpha})\bar{\alpha}(\lambda)=0$. The solution
is well known: $\bar{\alpha}(\lambda)=\frac \alpha {1-\frac
12l_f\alpha \ln \lambda}$, which fulfills the boundary condition:
$\bar{\alpha}(\lambda)|_{\lambda=1}=\alpha$. Note again we only
need the canonical EFT (QCD) coupling $g_s$ or $\alpha$. Thus for
massless case the effective coupling's scaling behavior is the
same as the running one. For massive case, the two differ in
scaling behavior. The effective charge should exhibit the correct
threshold behavior while the running one could not possess the
correct threshold behavior due to the absence of explicit mass
effects in its evolution equation. Thus the running coupling seems
to fail the decoupling theorem\cite{Appel}. This lead us to
section six.
\section{Appelquist-Carazzone decoupling theorem and underlying structures}
As is generally shown in subsection five B, in massive sectors of
an EFT, the agent constants (scales) generally appear in the
simple logarithmic function as parametrized in
Eq.(~\ref{largemass}). But such parametrization would lead to
failure of the decoupling theorem a la
Appelquist-Carazzone\cite{Appel} in the beta function or anomalous
dimensions\cite{EFT}. In this section we shall show how to resolve
the problem in the underlying structures' perspective.
\subsection{Decoupling and repartition}
First let us note that, in the underlying theory perspective, the
EFT parameters and the underlying parameters are grouped or
partitioned into two separate sets by the reference scale that
naturally appear in any physical processes (e.g., center energy in
a scattering ) according to the relative magnitudes: one effective
$[g]$, one underlying $\{\sigma\}$. The decoupling of a heavy
field in an EFT could only alter the EFT contents, not the
well-definedness of the whole theory. Physically, the heavy field
could no longer be excited in the EFT processes. Mathematically,
in the decoupling limit when an EFT mass scale ($M_H$) become
infinitely large for the rest EFT scales, the mass itself becomes
an underlying parameter. That means the decoupling induces a new
partition between the effective and underlying parameters, with
the union of the two sets kept conserved in the course of
decoupling:
\begin{eqnarray}
\label{repartition} &&([g];\{\sigma\})_{\text{EFT with}\ M_H}
 \Longrightarrow
([g]^{\prime};\{\sigma\}^{\prime})_{\text{EFT without}\ M_H};\\
\label{prime} && [g]\bigcup\{\sigma\}=
[g]^{\prime}\bigcup\{\sigma\}^{\prime},\ \ [g]^{\prime}\equiv
[g]/[M_H],\ \ \{\sigma\}^{\prime}\equiv \{\sigma\}\bigcup [M_H].
\end{eqnarray}This repartition yields a new EFT that differs from
the original one by a very massive field, hence a new set of agent
constants $\{{\bar c}^{\prime}\}$ is generated from this
repartition.

In the homogeneous rescaling of all dimensional parameters, this
repartition of parameters also leads to the repartition in the
scaling effects and hence different contents of EFT scaling and
anomalies are resulted. Thus the decoupling of a heavy EFT field
could be naturally formulated in the underlying structures'
scenario. Symbolically, this repartition means the following
rearrangement,
\begin{eqnarray}
\label{rearrange} &&\sum_\sigma d_\sigma\sigma
\partial_\sigma+ \sum_gd_g g\partial_g\Rightarrow
\sum_{\sigma}^{\prime} d_\sigma\sigma
\partial_\sigma+ \sum_g^{\prime}d_g g\partial_g.
\end{eqnarray} Here the summation is performed for the primed sets
defined above. Then the original underlying structures' decoupling
operation (low energy limit) $L_{\{\sigma\}}$ should be replaced
by that with respect to the new set
$\{\sigma\}^{\prime}(=\{\sigma,M_H\})$ when $M_H$ becomes much
larger than the rest of the EFT scales (especially than the center
energy):
$L_{\{\sigma\}^{\prime}}\equiv\lim_{\Lambda_{\{\sigma\}},M_H
\rightarrow \infty}$. This in turn leads to the new
parametrization in terms of agent constants,
\begin{eqnarray}
\label{oldlimit} &&L_{\{\sigma\}} \sum_\sigma d_\sigma\sigma
\partial_\sigma =\sum_{\bar c}d_{\bar c} \bar
{c}\partial_{\bar c},\\
&&L_{\{\sigma\}^{\prime}} \sum^{\prime}_\sigma d_\sigma\sigma
\partial_\sigma =\sum_{{\bar c}^{\prime}}d_{\bar c} {\bar
{c}}^{\prime}\partial_{{\bar c}^{\prime}}.\end{eqnarray} Then in
the theoretical limit that $M_H$ becomes extremely large
(comparable to the smallest underlying scale), a new set of agent
constants ($\{{\bar c}^{\prime} \}$) appear instead of the
original one ($\{\bar c\}$). That means the following
reorganization,\begin{eqnarray}
 \label{rearrange2}&&\sum_{\bar c}d_{\bar c} \bar
{c}\partial_{\bar c} + \sum_gd_g
g\partial_g\Longrightarrow\sum_{{\bar c}^{\prime}}d_{{\bar
c}^{\prime}} {\bar c}^{\prime}\partial_{{\bar c}^{\prime}} +
\sum_g^{\prime}d_g g\partial_g.
\end{eqnarray}

Thus in discussing the decoupling of heavy EFT particles, we
should first return to the formulation with the {\em explicit}
presence of the underlying parameters and the accompanying limit
operators, i.e., to work with $\{L_{\{\sigma\}^{\prime}}$, $\Gamma
([p],[g]^{\prime};\{\sigma\}^{\prime})\}$  and $\{L_{\{\sigma\}}$,
$\Gamma ([p],[g];\{\sigma\})\}$. Then from the configuration of
the parameters described in Eq.(~\ref{prime}), the decoupling of a
heavy EFT field could be formally be taken as the following two
operations: (i) repartitioning the EFT and underlying parameters;
(ii) taking the low energy limit with respect to the new
underlying parameters, including the heavy EFT field's mass. This
is the mathematical formulation of EFT field decoupling.
\subsection{Practical decoupling of EFT fields}
But in practical situation, the mass of the heavy particle is not
infinitely large comparing to the rest of the EFT scales. If the
energy or momentum transfer scale is not vanishingly small, then
the above mathematical decoupling limit does not apply in effect.
We must find alternative realization of decoupling in the
anomalous dimensions or in the 'running' of the EFT parameters
(C.f. Eqs.(~\ref{bacteria}) and (~\ref{running})).

For this purpose, let us review what we really mean by 'running'.
From Eq.(~\ref{preRGE1}) or (~\ref{preRGE2}) we see that, as the
underlying parameters or their agents scale homogeneously while
keeping the EFT parameters intact, this partial scaling just leads
to the 'running' or 'renormalization' of the latter. Thus the
partition determines the 'running' characteristics: what to scale
as underlying parameters versus what to scale as EFT parameters.
For the EFT parameters, the underlying ones and their agents vary
with the same scaling parameter $\lambda$: $\{\sigma \}
\Rightarrow \{\lambda^{d_\sigma }\sigma \}$ or $\{\bar c\}
\Rightarrow \{\lambda^{d_{\bar c}}\bar c \}$. That means once an
EFT scale becomes 'underlying' one, then it should vary with the
same scaling parameter as the underlying parameters or their
agents do: $\{\bar{c}^{\prime}\}=\{{\bar c}, M_H\} \Rightarrow
\{\lambda^{d_{\bar{c}^{\prime}}}\bar{c}^{\prime}\}=\{\lambda^{d_{\bar
c}}{\bar c},\lambda M_H\} $. Then we can understand the decoupling
of the heavy field with mass $M_H$ in EFT in the following sense:

(a). Before decoupling, namely, the momentum transfer $q^2$ or
other EFT scales is larger than $M^2_H$ (that is $M_H$ is an EFT
parameter), as the agent constants $\{\bar c\}$ vary with
$\lambda$ while keeping $M_H$ and other EFT parameters intact,
then the heavy field should generally contribute to the 'running'
of EFT parameters when the corresponding diagrams containing the
heavy line(s) due to the presence of the following variation
component as is generally shown in section five B,
\begin{eqnarray}
&&\delta_\lambda\left [\ln \frac{{\bar c}^2}{M_H^2}\right ]=\ln
\frac{[(1+\delta \lambda){\bar c}]^2}{M_H^2}-\ln \frac{{\bar
c}^2}{M_H^2}\neq0.
\end{eqnarray}

(b). When the momentum transfer scale $q$ is below the heavy field
threshold $q_{\text{thr}}=nM_H, n>1$, $M_H$ is no longer an EFT
parameter and scales in the same quantitative way as $\{\bar c\}$.
This prescription of scaling both $M_H$ and $\{\bar c\}$ in the
same way would yield no contribution to the 'running' of the rest
EFT parameters:\begin{eqnarray} &&\delta_\lambda\left [\ln
\frac{{\bar c}^2}{M_H^2}\right ]=\ln \frac{[(1+\delta
\lambda){\bar c}]^2}{[(1+\delta \lambda)M_H]^2}-\ln \frac{{\bar
c}^2}{M_H^2}=0.\end{eqnarray}That is to say, taking as part of the
underlying parameters or their agents, $M_H$ will cancel out the
some agents' contributions to the 'running' as they against each
other in the same logarithmic factor. It is then obvious that due
to such mechanisms the anomalous dimensions of the other EFT
parameters will alter by a finite amount ($\Delta{\bar \delta}_g$)
as the realization of heavy EFT field
decoupling:\begin{eqnarray}{\bar \delta}_g\Rightarrow {\bar
\delta}^{\prime}_g={\bar \delta}_g+\Delta{\bar
\delta}_g.\end{eqnarray}

This the resolution of the decoupling issue in the EFT parameters
in EFTs with underlying structures. It is not difficult to see
that this resolution is more close to the 'subtraction' solution
\cite{EFT}. The difference is that we switch the status of the
heavy masses rather than perform subtractions: take the heavy
masses as EFT or underlying parameters. Of course for the
'running' parameters we must employ matching conditions across the
decoupling thresholds\cite{matching}. This part is similar to
conventional approaches. This could also be realized by employing
appropriate boundary conditions that differ across the thresholds,
which should be equivalent to matching. We will further
demonstrate such understandings in various concrete applications
in the future. We hope our understanding could also be useful in
the heavy quark effects in deeply inelastic scattering\cite{PDF},
which is important both theoretically and phenomenologically. This
ends our main technical presentation.
\section{Discussion and summary}
We should emphasize that what we have done is to improve our
understanding of the UV ill-definedness of quantum field theories.
Based on the well known EFT philosophy, we show that the scaling
law analysis incorporating the postulated existence of underlying
short distance structures could lead to quite useful insights into
the renormalization issue though it is already a textbook stuff.
But the conventional machinery for renormalization is quite
complicated and not pedagogically easy to go with.

The underlying structures' scenario shows that the conventional
complications are just due to our oversimplified guess of the true
underlying structures implemented through various artificial
regularization methods, which usually yield us UV divergences that
should not have appeared in the true parametrization of the
underlying structures. Then many complications and superficial
unreasonableness associated with infinity manipulation could be
removed. Of course the low energy physics should be effectively
insensitive to the UV details, which is in fact the philosophical
basis for conventional renormalization programs. But if we take a
closer look at underlying structures scenario, we could see that a
number of important consequences could be readily obtained before
knowing the true underlying structures and without employing any
sort of infinity manipulation or subtraction. This situation, in
our opinion, is the real meaning of insensitivity. More
importantly, the essential indeterminacy in any EFT could be
easily read off from the unavailability of the underlying
parameters, which in turn implies that the agent constants must be
finally fixed or determined through physical boundaries or data,
which is also necessary in conventional renormalization
programs\cite{sterman}. One important result that might seem
trivial is that, though we do not the true values of these agent
constants, their positions could be determined through examining
the full scaling laws as explained by in section five. The
remaining task is to determine their physical values, which are
scheme and scale invariants in conventional terminology.

The scaling laws followed from the rescaling of both EFT and
underlying parameters take a number of different forms. The most
general one, described in Eq.(~\ref{scaling5}) or
(~\ref{scaling13}) is valid both at the level of individual graph
and at the complete vertex functions and their truncations. This
envisages a harmony notion in full scaling in EFTs with underlying
structures. This form or parametrization is not given effectively
in conventional approaches. The scaling of the agent constants
alone is shown to lead to the 'running' or finite
'renormalization' of the EFT parameters. The equations describing
this partial scaling arise as the natural consequences of taking
the decoupling limit of the underlying scale parameters, which
affords the RGE a physical interpretation. The most general form
of this decoupling theorem contain the St\"uckelberg-Peterman RGE
that describes also transformation across different
renormalization prescriptions. This byproduct could not be
obtained within any conventional generalization of Callan-Symanzik
equations. Moreover, we have shown that the decoupling of heavy
EFT fields could also be realized in a simple manner without
performing subtractions. It also follow from the insight provided
by deep exploration of the existence of the underlying structures
for EFTs.

Of course, we have here focused on more theoretical and logical
aspects of the underlying structures' scenario. These equations
and the byproducts thus obtained might have further consequences
and productive applications. In the meantime, we must stress that
our investigations here are just primary, further elaboration and
development are also necessary and worthwhile. We hope our primary
attempts here have demonstrated the nontriviality of the EFT
philosophy.

In summary, we derived several new versions of the scaling laws in
any EFT assuming the existence of nontrivial underlying structures
without the need of infinity manipulations and the associated
complications. The new equations are more general than the
conventional ones and provide more physical insights in our
understanding of renormalization and EFTs. The decoupling theorem
a la Applequist and Carazzone is still valid with the underlying
structures appropriately accounted. When the mass is not extremely
large, the decoupling could be implemented in a simple manner.
\section*{Acknowledgement}
I wish to thank W. Zhu for his helpful discussions on scalings.
This project is supported in part by the National Natural Science
Foundation under Grant No.s 10075020 and 10205004, and by the ECNU
renovation fund for young researcher under Grant No. 53200179.


\begin{references}
\bibitem{PWQFTbook}M.E. Peskin and D.V. Schroeder, {\em An
Introduction to Quantum Field Theory} (Addison-Wesley, MA, 1995),
Chap. 8; S. Weinberg, {\em The Quantum Theory of Fields}
(Cambridge University Press, Cambridge, 1995), Vol. I, Chap. 12.
\bibitem{BPHZ}Here we mean the BPHZ program, see, e.g., N.N.
Bogoliubov and D.V. Shirkov, {\em An Introduction to the Theory of
Quantized Fields} (4th edition, Wiley, NY, 1980).
\bibitem{SP}E.C.G. St\"ukelberg and A. Petermann, Helv. Phys.
Acta \textbf{26}, 499 (1953).
\bibitem{GML}M. Gell-Mann and F.E.
Low, Phys. Rev. \textbf{95}, 1300 (1954). \bibitem{HV} G. 't Hooft
and M. Veltman, Nucl. Phys. \textbf{B44}, 189 (1972).
\bibitem {WRGE}S. Weinberg, Phys. Rev. \textbf{D8}, 3479 (1973).
\bibitem{CS}C.G. Callan, Jr., Phys. Rev. \textbf{D2}, 1541 (1970); K.
Symanzik, Com. Math. Phys. \textbf{18}, 227 (1970).
\bibitem{PRD65yang}See, for example, J.-F. Yang and J.-H. Ruan,
Phys. Rev. \textbf{D65}, 125009 (2002)[arXiv:hep-ph/0201255] and
references therein.\bibitem{LENU}For examples in nucleon EFTs,
see, e.g., S.R. Beane, P.F. Bedaque, W.C. Haxton, D.R. Phillips
and M.J. Savage, in {\em Boris Ioffe Festschrift}, ed. M. Shifman
(World Scientific, Singapore, 2001)[arXiv:nucl-th/0008064].
\bibitem{scheme}P.M. Stevenson, Phys. Rev. \textbf{D23}, 2916
(1981); G. Grunberg, Phys. Rev. \textbf{D29}, 2315 (1984); S.J.
Brodsky, G.P. Lepage and P.B. Mackenzie, Phys. Rev. \textbf{D28},
228 (1983).
\bibitem{Wilson}K.G. Wilson, Phys. Rev. \textbf{179}, 1499
(1969).
\bibitem{Witten}E. Witten, Nucl. Phys. \textbf{B104},
445 (1976). Here by the decoupling notion we mean that there
should appear no infinities due to the postulated existence of the
underlying structures, not exactly the same as the original notion
defined by T. Appelquist and J. Carazzone in \cite{Appel}.
\bibitem{TOE}J.-F. Yang, arXiv:hep-th/9708104; invited
talk in: {\em Proceedings of the XIth International Conference
'Problems of Quantum Field Theory98'}, eds. B.M. Barbashov, G.V.
Efimov and A.V. Efremov (Publishing Department of JINR, Dubna,
1999), p.202[arXiv:hep-th/9901138]; arXiv:hep-th/9904055.
\bibitem{sterman}G. Sterman, {\em An Introduction to Quantum Field
Theory} (Cambridge University Press, Cambridge, England, 1993),
p.299.
\bibitem{maxwell}D.T. Barclay, C.J.
Maxwell and M.T. Reader, Phys. Rev. \textbf{D49}, 3480 (1994);
C.J. Maxwell and A. Mirjalili, Nucl. Phys. \textbf{B577}, 209
(2000); C.J. Maxwell and S.J. Burby, Nucl. Phys. \textbf{B609},
193 (2001).
\bibitem{Trace}S.L. Adler, J.C. Collins and A. Duncan, Phys. Rev.
\textbf{D15}, 1712 (1977); J.C. Collins, A. Duncan and S.D.
Joglekar, Phys. Rev. \textbf{D16}, 438 (1977).
\bibitem{LB}H.J. Lu and S.J. Brodsky, Phys. Rev. \textbf{D48},
3310 (1993).
\bibitem{Poleren}See, e.g., S. Willenbrock and G. Valencia, Phys.
Lett. \textbf{B259}, 373 (1991); R.G. Stuart, {\em ibid}.
\textbf{B262}, 113 (1991); A. Sirlin, Phys. Rev. Lett.
\textbf{67}, 2127 (1991); T. Bhattacharya and S. Willenbrock,
Phys. Rev. \textbf{D47}, 4022 (1993); M. Passera and A. Sirlin,
Phys. Rev. \textbf{D58}, 113010 (1998); B.A. Kniehl and A. Sirlin,
Phys. Rev. Lett. \textbf{81}, 1373 (1998).
\bibitem{Collins}J.C. Collins, Phys. Rev. \textbf{D10}, 1213
(1974).
\bibitem{Coq}R. Coquereaux, Ann. Phys. \textbf{125}, 401 (1980).
\bibitem{Coleman}S. Coleman, {\em Aspects of Symmetry}, (Cambridge
University Press, Cambridge, 1985), Chap. 3; J.-F. Yang,
arXiv:hep-ph/0212208.
\bibitem{opmixing}H. Kluberg-Stern and J.B. Zuber, Phys. Rev.
\textbf{D12}, 54 (1975); N.K. Nielsen, Nucl. Phys. \textbf{B97},
527, \textbf{B120}, 212 (1977).
\bibitem{Hooft}G. 't Hooft, in: {\it Recent Developments in Gauge
Theories}, eds. G. 't Hooft {\it et al} (Plenum, New York, 1980),
p.135.
\bibitem{Diffr}K. Symanzik, in: Jasic
(ed.), Lectures on High Energy Physics, Zagreb 1961 (Gordon and
Breach, NY, 1965); T.T. Wu, Phys. Rev. \textbf{125}, 1436 (1962);
J.G. Taylor, Suppl.al Nuovo Cimento \textbf{1}, 857 (1963); R.W.
Johnson, J. Math. Phys. \textbf {11}, 2161 (1970).
\bibitem{Caswell}W.E. Caswell and A.D. Kennedy, Phys. Rev.
\textbf{D25}, 392 (1982).
\bibitem{Chanow}M.S. Chanowitz and J. Ellis, Phys. Rev.
\textbf{D7}, 2490 (1973).
\bibitem{Appel}T. Appelquist and J. Carazzone, Phys.
Rev. \textbf{D11}, 2262 (1975).
\bibitem{EFT}S. Weinberg, Phys. Lett. \textbf{B91}, 51 (1980);
L.J. Hall, Nucl. Phys. \textbf{B178}, 75 (1981); B. Ovrut and H.
Schnitzer, Nucl. Phys. \textbf{B179}, 381 (1981), \textbf{B189},
509(1981);  W. Bernreuther and W. Wetzel, {\em ibid},
\textbf{B197}, 228 (1982); W. Wetzel, {\em ibid}, \textbf{B196},
259 (1982).
\bibitem{Weinb}S. Weinberg, Phys. Rev. \textbf{118}, 838 (1960).
\bibitem{AF}D.J. Gross and F. Wilczek, Phys. Rev. Lett.
\textbf{26}, 1343 (1973) ; H.D. Politzer, Phys. Rev. Lett.
\textbf{26}, 1346 (1973).
\bibitem{matching} G. Rodrigo and A. Santamaria, Phys.Lett.
\textbf{B313}, 441 (1993); K.G. Chetyrkin, B.A. Kniehl and M.
Steinhauser, Phys. Rev. Lett. \textbf{79}, 2184 (1997); G.
Rodrigo, A. Pich and A. Santamaria, Phys. Lett. \textbf{B424}, 367
(1998).
\bibitem{PDF}see, e.g., M.A. Aivazis, J.C. Collins, F.I. Olness
and W.K. Tung, Phys. Rev. \textbf{D50}, 3102 (1994); M. Buza, Y.
Matiounine, J. Smith, R. Migneron and W. L. van Neerven, Nucl.
Phys. \textbf{B472}, 611 (1996); R.S. Thorne and R.G. Roberts,
Phys. Rev. \textbf{D57}, 6871 (1998); M. Kr\"ammer, F.I. Olness
and D.E. Soper, {\em ibid}, \textbf{D62}, 096007 (2000).
\end{references}
\end{document}